\documentclass{bmcart}

\usepackage{amsthm,amsmath}
\usepackage[utf8]{inputenc} 

\usepackage{graphicx}

\startlocaldefs
\endlocaldefs

\begin{document}

\begin{frontmatter}

\begin{fmbox}
\dochead{Research}


\title{Improvement of response bandwidth and sensitivity of Rydberg Receiver using multi-channel excitations}

\author[
  addressref={aff1},                   
]{\fnm{Jinlian} \snm{Hu}}
\author[
  addressref={aff1,aff2},                   
  corref={aff1,aff2},                       
  email={ycjiao@sxu.edu.cn}   
]{\fnm{Yuechun} \snm{Jiao}}
\author[
  addressref={aff1},                   
]{\fnm{Yunhui} \snm{He}}
\author[
  addressref={aff1,aff2},                   
]{\fnm{Hao} \snm{Zhang}}
\author[
  addressref={aff1,aff2},                   
]{\fnm{Linjie} \snm{Zhang}}
\author[
  addressref={aff1,aff2},                   
  corref={aff1},                       
  email={zhaojm@sxu.edu.cn}   
]{\fnm{Jianming} \snm{Zhao}}
\author[
  addressref={aff1,aff2},
]{\fnm{Suotang} \snm{Jia}}


\address[id=aff1]{
  \orgdiv{State Key Laboratory of Quantum Optics and Quantum Optics Devices, Institute of Laser Spectroscopy},             
  \orgname{Shanxi University},          
  \city{Taiyuan 030006},                              
  \cny{P. R. China}                                    
}
\address[id=aff2]{%
  \orgdiv{Collaborative Innovation Center of Extreme Optics},
  \orgname{Shanxi University},
  \city{Taiyuan 030006},
  \cny{China}
}



\end{fmbox}


\begin{abstractbox}

\begin{abstract} 
We investigate the response bandwidth of a superheterodyne Rydberg receiver at a room-temperature vapor cell, and present an architecture of multi-channel lasers excitation to increase the response bandwidth and keep sensitivity, simultaneously. Two microwave fields, denoted as a local oscillator (LO) $E_{LO}$ and a signal field $E_{Sig}$, couple two Rydberg states transition of $|52D_{5/2}\rangle\to |53P_{3/2}\rangle$. In the presence of the LO field, the frequency difference between two fields can be read out as an intermediate frequency (IF) signal using Rydberg electromagnetically induced transparency (EIT) spectroscopy. The bandwidth of the Rydberg receiver is obtained by measuring the output power of IF signal versus the frequency difference between two fields. The bandwidth dependence on the Rabi frequency of excitation lasers is presented, which shows the bandwidth decrease with the probe Rabi frequency, while it is quadratic dependence on the coupling Rabi frequency. Meanwhile, we investigate the effect of probe laser waist on the bandwidth, showing that the bandwidth is inversely proportional to the laser waist. We achieve a maximum response bandwidth of the receiver about 6.8~MHz. Finally, we design an architecture of multi-channel lasers excitation for increasing the response and keeping the sensitivity, simultaneously. Our work has the potential to extend the applications of Rydberg atoms in communications. 
\end{abstract}



\end{abstractbox}
%

\end{frontmatter}




\section{Introduction}
Rydberg atom-based electrometry has made a significant progress~\cite{Fancher2021} in the last decade due to its property of large polarizabilities and microwave-transition dipole moments~\cite{Gallagher1994} and their advantages in the measurement of weak signals with high sensitivity, calibration-free, and intrinsic accuracy~\cite{Holloway2017,Holloway2018}. An optical Rydberg electromagnetically induced transparency  
(Rydberg-EIT) spectroscopy~\cite{Mohapatra2007} and Autler–Townes (AT) splitting~\cite{Tanasittikosol2011} has been used to detect the properties of electric field over a wide frequency range from DC to over 1 THz~\cite{Fan2015,Jiao2016,jiao2017,Jau2020,Sedlacek2012,Sedlacek13,SimonsMT2019}, as well as terahertz imaging~\cite{Wade2017} and sub-wavelength imaging of MW electric-field distributions~\cite{Holloway2014}. Recently, the use of Rydberg-atom superheterodyne technique~\cite{Jing2020,SimonsMT2019} by adding an auxiliary local oscillator (LO) field greatly improve the sensitivity of the Rydberg receiver to 55~nVcm$^{-1}$Hz$^{-1/2}$, and later to 30~nVcm$^{-1}$Hz$^{-1/2}$ by adding an optical pumping technique~\cite{Prajapati2021}. The superheterodyne technique also provides a method to determine the phase, frequency and angle-of-arrival of the signal field~\cite{RobinsonAmy2021}. 

In addition, the Rydberg atom-based receiver has an application in wireless communication, which has been used to receive baseband signals in a formation of amplitude modulation (AM)~\cite{Cox2018,Jiao2019,Song2019}, frequency modulation (FM)~\cite{Anderson2021}, and phase modulation (PM)~\cite{SimonsM2019,Holloway2019,Liu2022b}. However, temporal dynamics of Rydberg sensors are predominantly limited at sub-microsecond timescales~\cite{Meyer2018,sapiro2020TimeDependenceRydberg}, which limits its applications for detecting more complex waveforms and transmitting data with high rates. Some investigations aim to improve the data transfer capacity by employing multiple bands and multiple channels~\cite{Holloway2021, Zou2020}, multiple species~\cite{Holloway2019AIP} and spatially distributed atomic receiver array~\cite{Otto2021}. Further, the time-dependent response of Rydberg receiver to pulse-modulated electric fields is investigated under different laser conditions, which shows the origins of Rydberg-Atom electrometer transient response~\cite{bohaichuk2022OriginsRydbergAtomElectrometer,2022TVandvideo} by investigation of the rise and fall times for the receiver response to a transient signal, separately. The direct investigation of the response of a Rydberg atom-based superheterodyne receiver is not reported yet. All the aforementioned works use the EIT steady-state response so that the changes in the probe beam can only be measured after atoms reach steady-state. Recently, a new receiver concept based on spatiotemporal shaping of the probe beam within a Rydberg vapor cell is proposed to increase the response speed of Rydberg sensors to greater than 100~MHz~\cite{knarr2023SpatiotemporalMultiplexedRydberg}.

In this work, we investigate the response bandwidth of a Rydberg atom-based superheterodyne receiver.
The dependence of response bandwidth on the excitation laser conditions is investigated, including the probe laser Rabi frequency, the probe beam waist and the coupling laser Rabi frequency. The results show the bandwidth decreases with the probe Rabi frequency, while shows quadratic increasing with the coupling  Rabi frequency. In addition, we show the bandwidth is inversely proportional to the laser waist. A maximum response bandwidth of the receiver about 6.8~MHz is achieved by optimizing the laser parameters. The small probe laser Rabi frequency and small probe beam waist can increase the response bandwidth, while it causes a reduction of Rydberg atoms number, which lowers the sensitivity of receiver. Finally, we design a structure of two-channel lasers excitation, which realizes the increasing bandwidth, meanwhile guarantee the sensitivity. Our results show the improved response bandwidth is consistent with the reduction of beam waist, e.g. the reduction of transit time.

\section{Experimental setup}

\begin{figure}[htbp]
    \centering
    \includegraphics[width=1\textwidth]{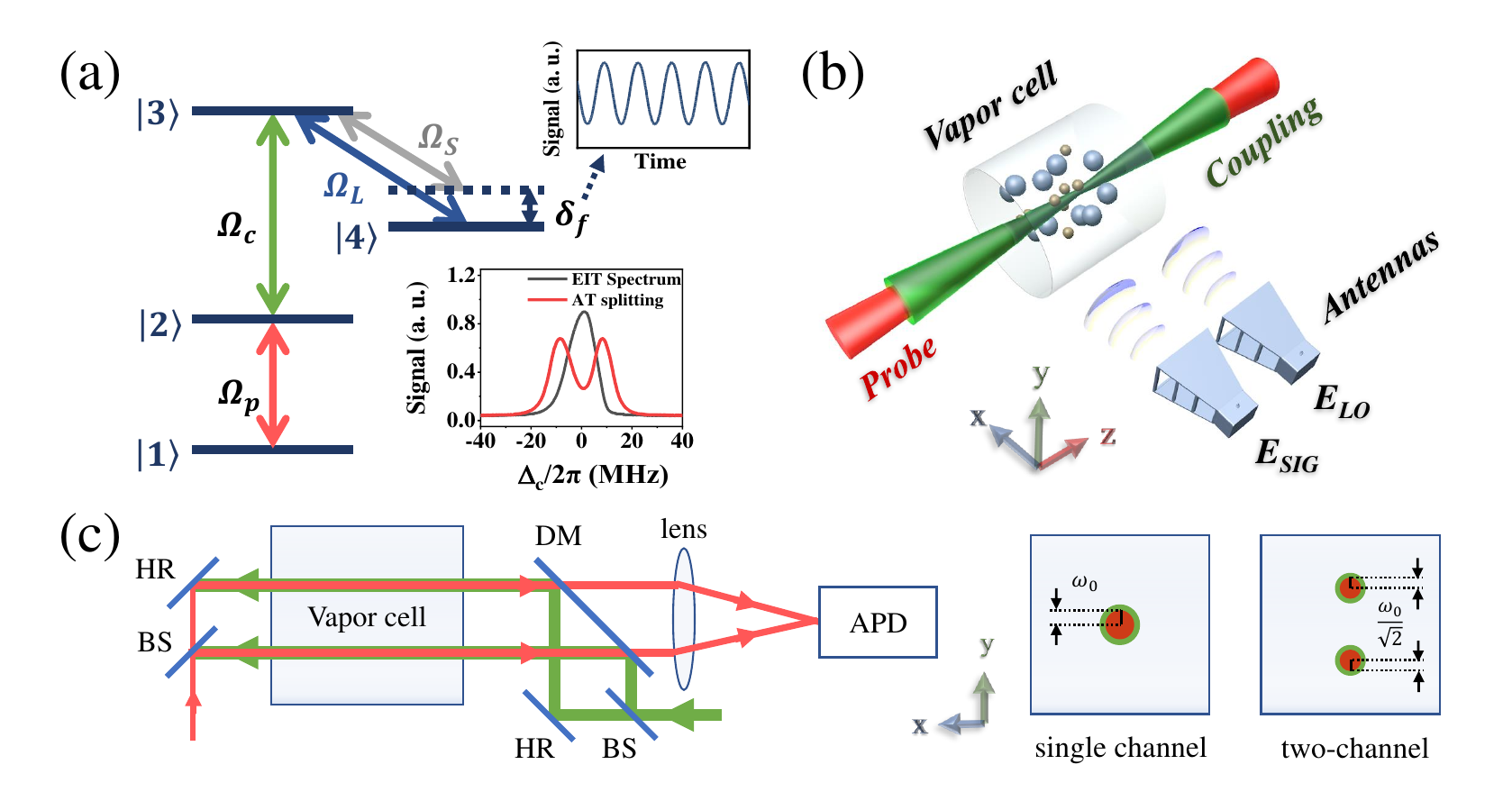}
    \caption{(a) Energy level diagram for the Rydberg EIT-AT system. A probe laser ($\Omega_p$) drives atoms from $|1\rangle$ to $|2\rangle$ and a coupling laser ($\Omega_c$) drives atoms from $|2\rangle$ to $|3\rangle$ transition. A local MW field ($\Omega_L$) resonantly couples the transition of $|3\rangle$ to $|4\rangle$, which causes the AT splitting of EIT spectrum, while the weak signal field ($\Omega_S$) has a detuning of $\delta_f$. In the presence of the LO field, the Rydberg atoms work as an atomic mixer to down-convert the signal field to an intermediate frequency with a frequency of $\delta_f$, which causes an oscillation in the transmission of the probe laser. (b) Sketch of the experimental setup for single channel Rydberg receiver. A coupling laser and a probe laser counter-propagate through a cylindrical room-temperature cesium cell. The transmission of the probe laser is detected by an avalanche photodiode (APD). Two RF fields, noted as the LO field and the signal field ($E_{LO}$ and  $E_{SIG}$), are applied transversely to the laser beams propagating through the vapor cell by two identical horn antennas. (c) Diagram of two-channel Rydberg receiver for increasing the response bandwidth. Both the probe and coupling lasers are split into two beams using a 
    50/50 beam splitter (BS) and counter-propagate through the cell. One channel is right above the other. The beam waists of each probe and coupling laser are set to $\omega_p/\sqrt{2}$ and $\omega_c/\sqrt{2}$. After propagating through the cell, two separated probe lasers pass through a dichroic mirror (DM) and are merged into a detector.}
    \label{figure1}
\end{figure}

We perform the experiment in a cylindrical cesium room-temperature vapor cell with $\phi$ 25 mm $\times$ 20 mm. A relevant four-level Rydberg EIT-AT scheme with $^{133}$Cs atoms and a diagram of setup for single-channel Rydberg receiver are illustrated in Figs.~\ref{figure1} (a) and (b). A green Rydberg coupling laser ($\lambda_c$ = 510~nm) and a near-infrared probe laser ($\lambda_p$ = 852~nm) counter-propagate through the cell, where the probe laser drives the transition of $|1\rangle\to |2\rangle$ ($|6S_{1/2}, F = 4\rangle\to |6P_{3/2}, F^\prime = 5\rangle$) with Rabi frequency $\Omega_p$ and the coupling laser further excites the atoms to the Rydberg state $|3\rangle$ ($|52D_{5/2}\rangle$) with Rabi frequency $\Omega_c$, thus establishing the EIT spectroscopy to enhance the probe transmission. The EIT signal is detected by measuring the transmission of the probe laser with an avalanche photodiode (APD). The probe and coupling lasers keep co-linear polarization along the y-axis and their $1/e^2$ beam waist $\omega_p$ and $\omega_c$ are varied. Two MW fields with co-linear polarization along the y-axis, denoted as a strong local oscillator (LO) field $E_{LO}$ and a weak signal field $E_{Sig}$, are emitted from two identical horn antennas (A-info LB-20180-SF) and incident to the Rydberg system simultaneously. In our experiment, we fix the $E_{LO}$ = 5.5 mV/cm ($\Omega_L$ = 2$\pi \times$ 12.5 MHz) and $E_{Sig}$ = 1.2 mV/cm ($\Omega_S$ = 2$\pi \times$ 2.27 MHz) for the experiments. The choose of LO MW filed strength is used to obtain the maximum sensitivity, and its value could also have an effect on the response time~\cite{Meyer2018}. The LO field is resonant with the transition of two adjacent Rydberg states $|3\rangle\to |4\rangle$ ($|52D_{5/2}\rangle\to |53P_{3/2}\rangle$) with frequency of 5.038 GHz, leading to AT splitting of EIT spectrum, which is proportional to the Rabi frequency of LO field, $\Omega_L$. While the weak signal field couples two Rydberg states with a detuning of $\delta_f$ and Rabi frequency of $\Omega_S$. In the presence of the LO field, the Rydberg atoms work as an atomic mixer to down-convert the signal field to an intermediate frequency with frequency of $\delta_f$, which causes an oscillation in the transmission of the probe laser, shown in Fig.~\ref{figure1} (a), where the oscillation signal is interrogated by a spectrum analyzer (ROHDE $\&$ SCHWARZ FSVA3013). Fig.~\ref{figure1}~(c) shows the diagram of a 2-channel  version of a Rydberg receiver for increasing the response bandwidth. One channel is right above the other, which make sure the oscillation signals in-phase detected by the PD. The distance between two channels is 10~mm to exclude the exchange of Rydberg atoms between two channels. Both the probe and coupling lasers are split into two beams using 50/50 beam splitter (BS) and counter-propagate through the cell, and the beam waists of probe and coupling laser are set to $\omega_p/\sqrt{2}$ and $\omega_c/\sqrt{2}$, such keeping the probe (coupling) Rabi frequency of each channel equal to the probe (coupling) Rabi frequency in single channel, while the only difference is that the excitation volume of each channel is equal  to the half of 
single channel. After propagating through the cell, two separated probe lasers pass through a dichroic mirror (DM) and are merged into a detector. 

\section{Results and discussions}
\begin{figure}[htbp]
    \centering
    \includegraphics[width=1\textwidth]{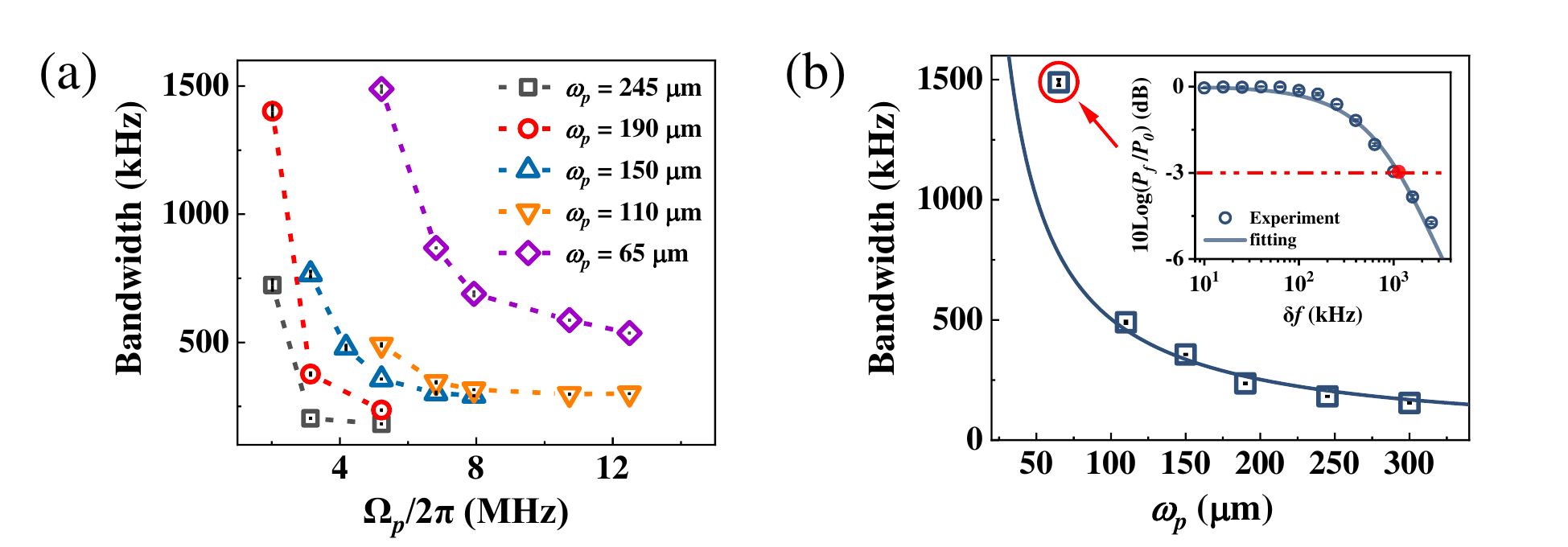}
    \caption{ (a) Measured response bandwidths versus the probe laser Rabi frequencies ($\Omega_p$) for five different sets of probe beam waists. The coupling laser beam waists were adjusted during the experiments to maintain their slightly larger than the probe, and its Rabi frequency is fixed to $\Omega_c$ = 2$\pi \times$ 4.55 MHz. The response bandwidth decreases with the probe laser Rabi frequency for each probe beam waist. (b) The dependence of response bandwidths on the probe beam waists with fixing probe laser Rabi frequency and coupling laser Rabi frequency $\Omega_p$ = 2$\pi \times$ 5.22 MHz, $\Omega_c$ = 2$\pi \times$ 4.55 MHz. The atomic response bandwidth decreases with the probe beam waist. The solid line shows the fitting result. Inset: The output power as a function of $\delta_f$ obtained with a spectrum analyzer at the probe beam waist $\omega_p$ = 65 $\mu m$. The 3~dB bandwidth is about 1.5 MHz. The error bars are standard deviations of 10 measurements for all data sets.}
    \label{figure2}
\end{figure}

For extraction of the response bandwidth, we collect a series of sinusoidal oscillation signals such as in Fig.~\ref{figure1} (a) and measure their output power using the spectrum analyzer as a function of $\delta_f$ by changing the signal field frequency with fixing LO field frequency. The response bandwidth of the Rydberg receiver is defined as the frequency corresponding to the output of the spectrum analyzer attenuating 3dB of its power. The inset in Fig.~\ref{figure2} (b) shows an example that the output power attenuates with increasing the $\delta_f$, corresponding to the 3~dB bandwidth about 1.5~MHz.

The response bandwidths of Rydberg system consist of the time of reaching steady state and the transit time~\cite{bohaichuk2022OriginsRydbergAtomElectrometer}, which are related to Rabi frequency of excitation lasers and the excitation volume. When we study the steady-state part, the transit time is fixed, while when we study the transit time, the time of build steady-state is fixed. In Fig.~\ref{figure2} (a), we demonstrate the effects of probe Rabi frequency ($\Omega_p$) on the response bandwidths of the superheterodyne receiver for five different sets of probe beam waists. Each data set in Fig.~\ref{figure2} (a) is taken with fixing the coupling laser Rabi frequency $\Omega_c$ = 2$\pi \times$ 4.55 MHz and the coupling laser waist slightly larger than the probe laser waist. For giving more details of experimental parameters, as an example for $\omega_p$ = 65 $\mu m$, the probe Rabi frequency is varied from 2$\pi \times$ 5.22 MHz to 2$\pi \times$ 12.49 MHz, corresponding to probe power varied from 0.22 $\mu$W to 1.26 $\mu$W, while the coupling laser beam waist is $\omega_c$ = 80 $\mu m$ and its power is 10 mW. All results show that the response bandwidths decrease with $\Omega_p$ at certain beam waist, which is consistent with the previous work that the probe Rabi frequency lows to 2$\pi \times$ 0.65MHz ~\cite{bohaichuk2022OriginsRydbergAtomElectrometer}. Therefore, it is advantageous to use lower $\Omega_p$ to get a fast response for fast-timing applications. This is because the steady state is rapidly built at lower $\Omega_p$, in response to effective alteration of the EIT by the MW field. The results also show the bandwidths are disparate for different probe beam waists with the same $\Omega_p$.

Then, with fixing the probe Rabi frequency and coupling Rabi frequency to $\Omega_p$ = 2$\pi \times$ 5.22 MHz (the probe laser power is varied from 0.22 $\mu$W to 4.70 $\mu$W to keep its Rabi frequency constant) and $\Omega_c$ = 2$\pi \times$ 4.55 MHz (the coupling laser power is varied from 10.0 mW to 165.0 mW to keep its Rabi frequency constant), we investigate the dependence of response bandwidths on the probe beam waists by varying the waist of probe beam from 65 $\mu m$ to 300 $\mu m$. The results in Fig.~\ref{figure2}(b) show the response bandwidth depends strongly on the probe beam waist. The varied beam waist changes the transit time, which is used to describe the time that the Rydberg atoms move in and out the interaction volume~\cite{2022TVandvideo}. The transit time can be expressed as:
\begin{equation}\label{transit time}
    \tau_{t}=\frac{2\cdot \omega}{v},
\end{equation}
where $\omega$ is the beam waist, $v$ is the average velocity of Cs atoms, which is 218 m/s at room temperature. Using above equation, we get the transit time is 0.59 $\mu$s and 2.75 $\mu$s for probe beam waist of 65 $\mu m$ and 300 $\mu m$, respectively. We simply fit the experimental data using $\frac{A}{\tau_{t}}$, yielding A = 0.46 that account for other factors contributing to the bandwidth, e.g. the time of reaching steady-state, which is related to Rabi frequency of probe and coupling lasers.
The solid line shows the fitting result, which agrees well with the data, except for the 65 $\mu m$ width. This is because the strong divergence of the probe laser makes the average beam waist through the cell be about 114 $\mu m$, which is larger than 65 $\mu m$, so the effective Rabi frequency is lower than estimated. As the above study showed, the lower Rabi frequency gets the higher response, so the actual measured bandwidth is larger than the fitting point. Our investigation shows that the response bandwidth is inversely proportional to the beam waist $\omega$. 

\begin{figure}[htbp]
    \centering
    \includegraphics[width=0.5\textwidth]{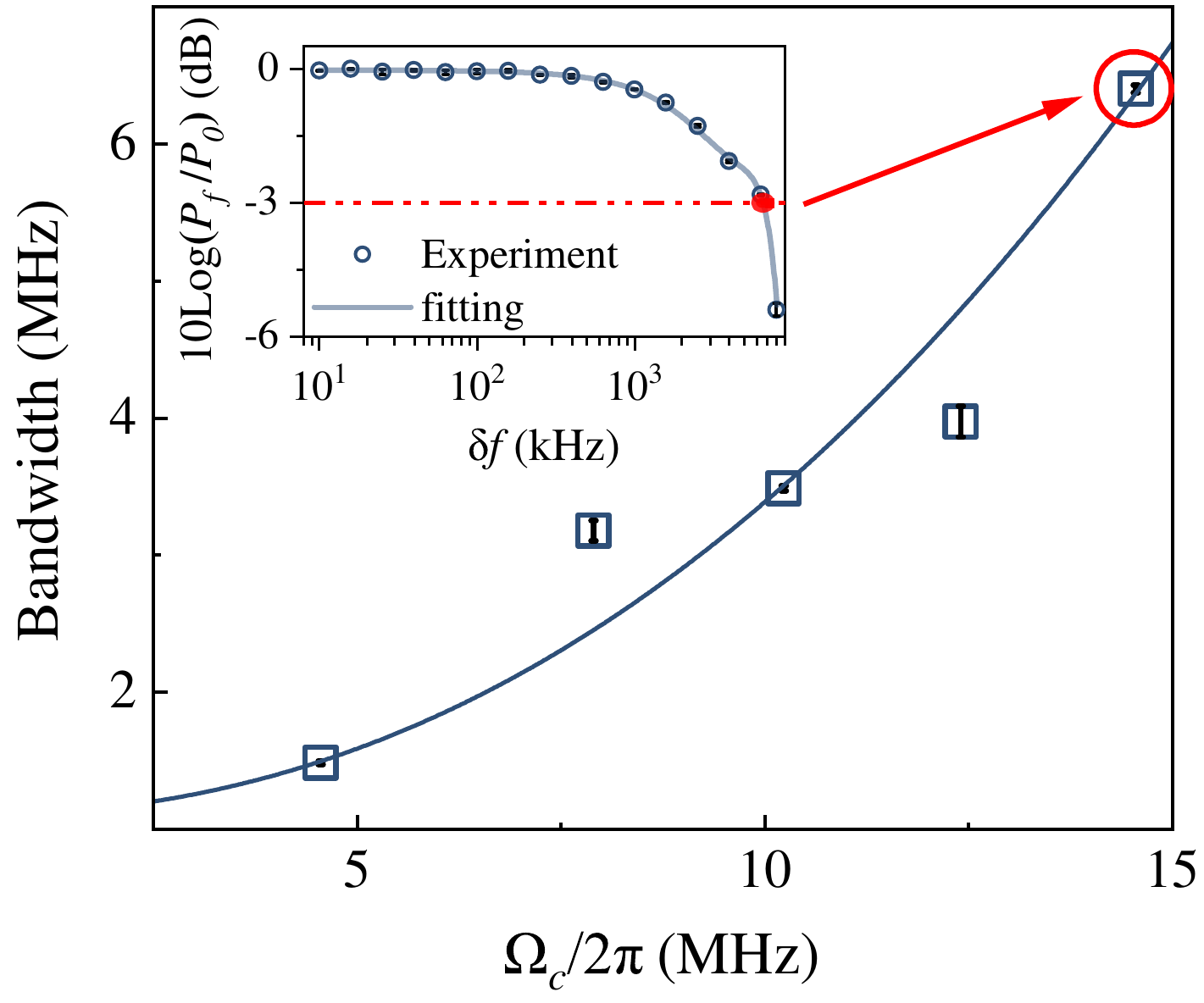}
    \caption{The response bandwidths of the Rydberg system versus the coupling Rabi frequencies ($\Omega_c$) with fixed $\omega_p$ = 65 $\mu m$ and $\Omega_p$ = 2$\pi \times$ 5.22 MHz, which shows increasing with the coupling laser Rabi frequency. The solid line shows the fitting result. Inset: The output power as a function of $\delta_f$ obtained with a spectrum analyzer at coupling Rabi frequency $\Omega_c$ = 2$\pi \times$ 14.55 MHz. The 3~dB bandwidth is about 6.8 MHz. The error bars are standard deviations of 10 measurements for all data sets.}
    \label{figure3}
\end{figure}

Next, we measure the response bandwidths of the Rydberg system versus the coupling Rabi frequencies ($\Omega_c$) with fixed $\omega_p$ = 65 $\mu m$, $\omega_c$ = 80 $\mu m$ and $\Omega_p$ = 2$\pi \times$ 5.22 MHz. The results are presented in Fig.~\ref{figure3}, which shows the response bandwidth increases with the $\Omega_c$ varied from 2$\pi \times$ 4.55 MHz to 2$\pi \times$ 14.55 MHz, corresponding coupling laser power varied from 10.0 mW to 102.2 mW. The higher coupling Rabi frequencies can accelerate reach the steady state, such getting a faster response. The bandwidth could be continued to scale if we have stronger coupling laser power before it is saturated.  The inserted subplot shows an example for the output power as a function of $\delta_f$ obtained with a spectrum analyzer at coupling Rabi frequency $\Omega_c$ = 2$\pi \times$ 14.55 MHz. The 3~dB bandwidth is 6.8 MHz. We fit the data using $A\cdot x^a + B$, yielding a = 2.2, which shows a roughly quadratic relationship between response bandwidths and coupling Rabi frequencies. The solid line shows the fitting result. 

As the LO field breaks the EIT and the signal electric field much weaker than the coupling laser, the response speed of Rydberg atoms is mainly determined by the timescale of establishment EIT. So we consider a three-level system to understand the dependence of response bandwidths on the coupling Rabi frequencies, we theoretically analyze the relationship between steady-state time $\tau_s$ and coupling laser Rabi frequency $\Omega_c$. The dynamics of the system are governed by Lindblad master equation ($\hbar=1$)
\begin{equation}\label{master}
    \dot{\rho}(t)=-i[H,\rho]+\mathcal{L}(\rho),
\end{equation}
where $\rho$ is the density matrix, $H$ is the Hamiltonian of three-level atomic system, and $\mathcal{L}(\rho)$ is the Lindblad operator describing the dissipation. We assume the initial condition of the system is in the ground state $|1\rangle$, i.e. $\rho_{11} (0)=1, \rho_{22} (0)=0, \rho_{33} (0)=0$. Here we make the approximation that the contribution of the upper transition dipole amplitude $\rho_{23}$ is neglected. Since the energy levels 2 and 3 are barely populated by atoms and the coherence is neglected, so $\rho_{23}$ = 0. We also assume $\dot{\rho}_{13}=0$, which is mainly due to the fast-responding spin wave between levels 1 and 3, whose contribution is negligible when the EIT reaches the steady-state. By solving Eq.~(\ref{master}), we can obtain the time evolution of the coherence between levels 1 and 2,
\begin{equation}\label{drho12}
    \dot{\rho}_{12}=-(i\Delta_p+\gamma_{12})\rho_{12}+i\frac{\Omega_p}{2}-\frac{\Omega_c^2}{4[i(\Delta_p+\Delta_c)+\gamma_{13}]}\rho_{12},
\end{equation}
where $\gamma_{ij}=(\Gamma_i+\Gamma_j)/2$, $\Gamma_{i(j)}$ is the spontanous decay rate. $\Delta_p$ and $\Delta_c$ are the detunings of the probe laser and the coupling laser respectively, $\Omega_p$ and $\Omega_c$ are the Rabi frequencies of the probe laser and the coupling laser. We derive the analytical solution using Eq.~(\ref{drho12})
\begin{equation}\label{rho12}
    \rho_{12}(t)=\frac{2\Omega_p \gamma_{12}+C_1 e^{-\frac{t\sigma_1}{4((\Delta_p+\Delta_c )-i\gamma_{13} )} }+2i\Omega_p (\Delta_p +\Delta_c)}{\sigma_1} ,
\end{equation}
    where $\sigma_1=4\gamma_{12}(\Delta_p+\Delta_c)+4\Delta_p\gamma_{13}-4i\gamma_{12} \gamma_{13}-i\Omega_c^2+4i\Delta_p(\Delta_p+\Delta_c) $ and $C_1$ is the amplitude of the coherence of the probe field, can be any value by changing the experimental conditions. As the Im$(\rho_{12})$ denotes the absorption of the probe laser, we find only the exponential part evolves with time,
\begin{equation}\label{e(t)}
    e^{-\beta t}=e^{-t\frac{4((\Delta_p+\Delta_c )^2 +\gamma_{13}^2) \gamma_{12}+\gamma_{13} \Omega_c^2}{4((\Delta_p+\Delta_c )^2 +\gamma_{13}^2)}} ,
\end{equation}
 Then steady-state time $\tau_{s}$ is
\begin{equation}\label{t}
    \tau_s=\frac{1}{\beta}=\frac{4((\Delta_p+\Delta_c )^2 +\gamma_{13}^2)}{4((\Delta_p+\Delta_c )^2 +\gamma_{13}^2) \gamma_{12}+\gamma_{13} \Omega_c^2} ,
\end{equation}
The response bandwidth $B$ is the inverse of $\tau_s$, so we can get $B\propto \frac{1}{\tau_s} \propto \Omega_c^2$ with $\Delta_p$ = $\Delta_c$ = 0, which also shows the quadratic relationship. Note that our theoretical analysis is used to obtain the dependence of the bandwidth on the coupling Rabi frequency in the regime of $\Omega_c > \Omega_p$, may not suit to  $\Omega_c < \Omega_p$.

\begin{figure}[htbp]
    \centering
    \includegraphics[width=0.8\textwidth]{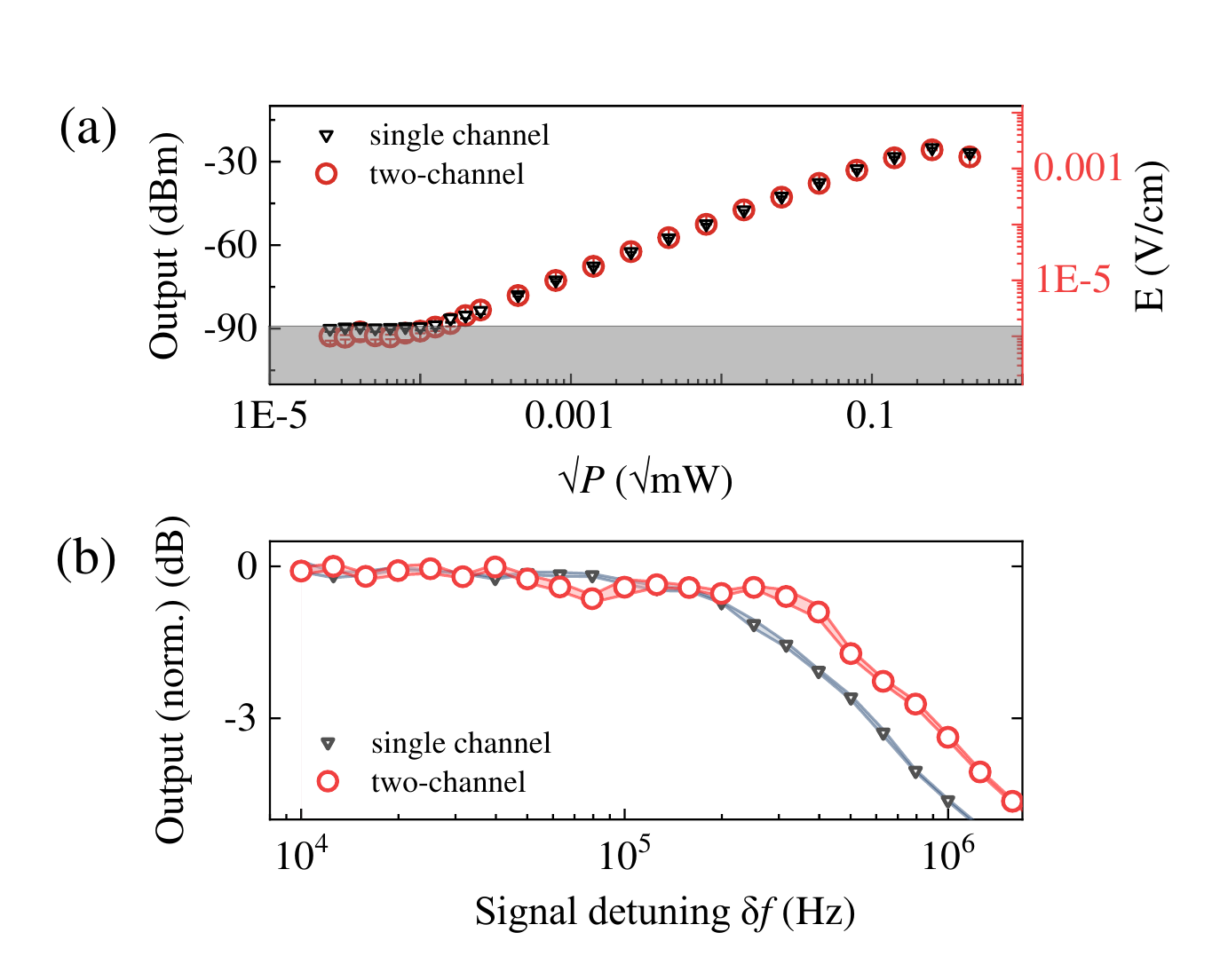}
    \caption{ (a) Comparison of sensitivity between two-channel scheme (red hollow circles) and single channel scheme (black triangles). Each probe (coupling) Rabi frequency of two-channel structure is equal to the single channel, and the total interaction volume of the two-channel structure is same with that of single channel. The results show the sensitivity about 477~nVcm$^{-1}$Hz$^{-1/2}$ for the single channel and the sensitivity about 468~nVcm$^{-1}$Hz$^{-1/2}$ for the two-channel, which are almost same for both scheme. (b)  Comparison of response bandwidth between two-channel scheme (red hollow circles) and single channel scheme (black triangles). The results show the two-channel scheme ( red hollow circles, 872~kHz) has an improvement compared with the single channel scheme (black triangles,578~kHz). The improved factor is 872/578 $\approx$ 1.5, which is consistent with the reduction of beam waist by $\sqrt{2}$, e.g. the reduction of transit time. The error bars are standard deviations of 10 measurements for all data sets.}
    \label{figure4}
\end{figure}
The above investigations reveal we get fast response for smaller probe Rabi frequency and smaller probe beam waist. However, this will cause a reduction of Rydberg atoms number, further lowering the sensitivity of receiver. In order to keep faster response and high sensitivity at the same time, we design a structure of two-channel lasers excitation, see Fig.~1(c). In two-channel structure, the probe (coupling) Rabi frequency of each channel is same and equal to the single channel Rabi frequency. 
The beam waists of probe and coupling lasers in each channel are set to $\omega_p/\sqrt{2}$ and $\omega_c/\sqrt{2}$, such that the laser Rabi frequency and excitation volume of two-channel design are same as the single channel. In our experiment, we set $\Omega_p$ = 2$\pi \times$ 7.55~MHz and $\Omega_c$ = 2$\pi \times$ 4.55~MHz for both schemes, where $\omega_p$ = 92~$\mu m$, $\omega_c$ = 131~$\mu m$, probe power 0.92 $\mu$W and coupling power 26.8 mW for the single channel scheme, and $\omega_p$ = 65~$\mu m$, $\omega_c$ = 93~$\mu m$, probe power 0.46 $\mu$W and coupling power 13.4 mW for the two-channel scheme. Fig.~\ref{figure4}(a) and Fig.~\ref{figure4}(b) show the comparison of the sensitivity and response bandwidth of two-channel scheme with the single channel scheme, respectively. In Fig.~\ref{figure4}(a), we test the detectable signal as a function of the MW power with $\delta_f$ = 10~kHz for two-channel scheme (red hollow circles) and single-channel scheme (black triangles). After calibration~\cite{hu2022ContinuouslyTunableRadio}, we get the sensitivity about 477~nVcm$^{-1}$Hz$^{-1/2}$ for the single channel and the sensitivity about 468~nVcm$^{-1}$Hz$^{-1/2}$ for the two-channel, which shows the sensitivity for both scheme are almost same. In Fig.~\ref{figure4}(b), the measurement of bandwidth shows the single channel scheme (black triangles) is about 578~kHz, while the two-channel scheme has been improved to 872~kHz. The improved factor is 872/578 $\approx$ 1.5, which is consistent with the reduction of beam waist by $\sqrt{2}$, e.g. the reduction of transit time.

\section{Conclusions}
We have investigated the response bandwidth of a Rydberg atom-based superheterodyne receiver system on the Rabi frequency of excitation lasers and the excitation volume. We show the response bandwidth decreases with the probe Rabi frequency, while it shows a quadratic relationship with the coupling Rabi frequency. The investigation of dependence of the response bandwidth on the laser waist shows it is inversely proportional to the laser waist. We achieved a response bandwidth of the receiver about 6.8~MHz by optimizing the laser parameters. In addition, we give a theory to describe the experimental results, which shows good agreement. Finally, we designed a two-channel lasers excitation structure to get faster response and keep high sensitivity simultaneously. The improved response bandwidth is consistent with the reduction of beam waist, e.g. the reduction of transit time. Our work has the potential to extend the applications of Rydberg atoms in communications.


\begin{backmatter}

\section*{Acknowledgements}

\section*{Funding}
This work is supported by the National Key Research and Development Program of China (2022YFA1404003); National Natural Science Foundation of China (Grant Nos. 61835007, 12120101004, 62175136, 12241408); the Scientific Cooperation Exchanges Project of Shanxi province (Grant No. 202104041101015); Program for Changjiang Scholars and Innovative Research Team in University (IRT 17R70); 1331 project of Shanxi province;

\section*{Abbreviations}
local oscillator (LO), intermediate frequency (IF), electromagnetically induced transparency (EIT).

\section*{Availability of data and materials}
The data that support the findings of this study are available
from the corresponding author upon reasonable request.

\section*{Competing interests}
The authors declare that they have no competing interests.

\section*{Authors' contributions}
J.H. and Y.J. did the experiment. Y.H. calculated the theory. All authors contributed to all the parts of the paper. All authors read and approved the final manuscript.


\bibliographystyle{bmc-mathphys} 
\bibliography{bmc_article}      



\end{backmatter}
\end{document}